\documentclass[aps,prc,twocolumn,showpacs,floatfix,nofootinbib,preprintnumbers,groupaddress,amsmath,amssymb]{revtex4-1}
\usepackage[T1]{fontenc}
\usepackage[utf8]{inputenc}
\usepackage{graphicx}
\usepackage{epsfig}
\usepackage{bm}
\usepackage{color}
\usepackage{float}
\usepackage{dcolumn}
\usepackage{txfonts}

\graphicspath{ {images/} }

\begin{document}

\title{Evidence for the virtual $\beta$-$\gamma$ transition in $^{59}$Ni decay}

\author{M.~Pf\"utzner}

\author{K. Pachucki}

\author{J. Żylicz}

\affiliation{Faculty of Physics, University of Warsaw, ul. Pasteura 5, 02-093 Warsaw, Poland}


\begin{abstract}
A novel theory of radiative electron capture in second forbidden
non-unique transition, is applied to
reanalyze experimental data obtained previously for the decay of $^{59}$Ni.
The measured gamma spectrum is shown to be distorted at the high-energy end, presenting
the first direct evidence for the virtual beta-gamma transition through the excited
state in $^{59}$Co. The complete theoretical predictions, including both the direct and the virtual
decay channels, as well as the interference between them, reproduce
very well the experimental spectrum. The virtual nuclear component amounts to about 4\%
of the total gamma intensity.
\end{abstract}


\maketitle

\section{Introduction}
Orbital electron capture by an atomic nucleus (EC)
is one of basic nuclear transitions governed by weak interactions \cite{Bambynek}.
Although known and investigated for a very long, it is still a subject
of interest in the various fields of fundamental research.
Examples include neutrino-less double-EC decay \cite{SujkowskiWycech},
search for sterile neutrinos \cite{Filianin}
and testing the Lorentz invariance \cite{Vos}.
A particularly interesting aspect of EC decay, not yet fully explored,
is the coupling of electronic and nuclear degrees of freedom, as
shown in the decay of highly charged ions \cite{Litvinov}
or in the searched nuclear excitations by electron capture \cite{Palffy}.
Here, we address one such peculiar phenomenon where to account for radiation
emitted during the EC decay, both the electronic and nuclear contributions have to be
summed coherently. In addition, each contribution depends on the electromagnetic gauge,
and in the so-called length gauge employed here, the nuclear radiation is suppressed.

According to quantum mechanics, the evolution of a
physical system can be described as proceeding through all possible paths,
including those which may temporarily violate the law of energy conservation.
The occurrence of such special paths, involving virtual intermediate states, becomes important
(and may be observable) when all other possibilities are not allowed or are highly forbidden.
An example of such a situation is the nuclear beta decay in which the direct transition
between the initial and the final state is strongly hindered due to angular momentum and/or parity
selection rules, while no other decay channels are present within the available energy window.
Then, as noted already long ago by Longmire \cite{Longmire}, an alternative decay
path may involve a state in the daughter nucleus having energy higher than the initial state.
If beta transition to such a state is allowed, a combination of this transition with
emission of a gamma-ray may compete with the direct decay. Rose, Perrin, and Foldy discussed
such a virtual beta-gamma emission in more detail and noted that $^{59}$Ni
is a promising candidate to observe this phenomenon \cite{Rose}. The ground state
of $^{59}$Ni decays to the ground state of $^{59}$Co with the
half-life of $(1.08 \pm 0.13) \times 10^5$~y \cite{T1/2},
predominantly by electron capture with a tiny contribution of $\beta ^+$ \cite{NDS}.
The direct transition
between these two states involves the change of spin by two units and no change of parity, thus
it is classified as a second forbidden non-unique (\emph{2nu}) transition \cite{Bambynek}.
However, in $^{59}$Co there exists a state located about 26~keV above the ground state
of $^{59}$Ni, offering the possibility of the virtual "detour" transition by a combination
of allowed (Fermi+Gamow-Teller) electron capture, followed by E2 gamma emission.
Both decay paths and the current knowledge on the involved states are shown in Figure 1.
The direct \emph{2nu} transition may also proceed with the emission of a photon. Such a
second-order, radiative electron capture (REC) occurs with the probability of about $10^{-4}$
with respect to the ordinary, non-radiative capture. The dominant source of this radiation is the
internal bremsstrahlung (IB) of the captured electron \cite{Bambynek}.
The REC energy spectrum is continuous, because of energy sharing between three
bodies in the final state. The maximum photon energy equals the decay energy $Q_{EC}$
minus the binding energy of the captured electron in the daughter atom. In case of the K-capture
in $^{59}$Ni, the REC spectrum extends up to $Q_{K}=1064.8$~keV. The virtual
"detour" transitions were predicted to modify the REC spectrum,
mainly at the high energy end \cite{Rose}. The more extended calculation of this
process, including interference effects, was done by Lassila \cite{Lassila}.

\begin{figure}[htb]
\includegraphics[width=0.9\columnwidth]{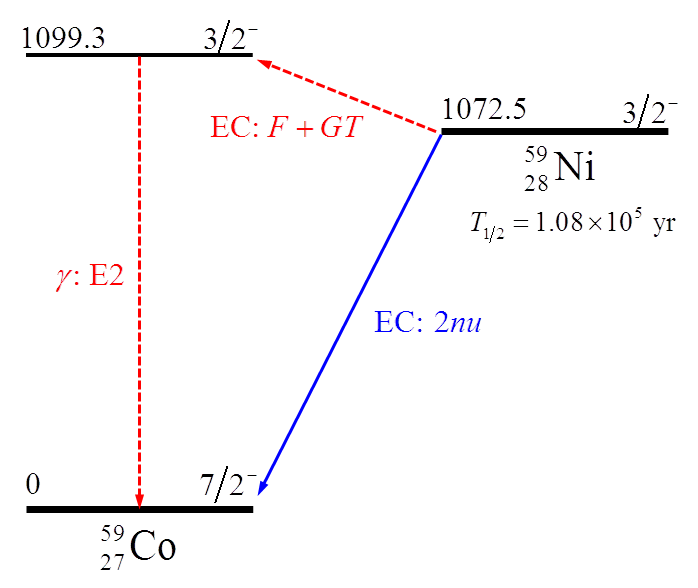}
\caption[T]{\label{fig:fig1}
(Color online) Decay scheme of $^{59}$Ni and relevant states in $^{59}$Co \cite{NDS}.
The energy levels are labeled by energy in keV, relative to the ground state
of $^{59}$Co and by spin and parity values. The direct decay of $^{59}$Ni proceeds
by second forbidden non-unique electron capture to the ground state of $^{59}$Co (solid arrow).
The virtual "detour" transition goes by allowed (Fermi and Gamow-Teller) electron
capture to the excited state of $^{59}$Co combined with gamma E2 deexcitation
(dashed arrows).
}
\end{figure}

The REC of $^{59}$Ni has already been studied several times. Except for the very first
attempt \cite{Saraf}, all later experiments were motivated by the search for
the virtual beta-gamma transitions \cite{Schmorak, Berenyi, Janas}.
Only Ref. \cite{Schmorak} claimed an observation of these transitions.
However, as we argue below, results of these papers could not have been
conclusive because the applied model of the REC process was either inaccurate or erroneous.
We note in addition, that the virtual transitions were also considered for
the case of $\beta^-$ decay \cite{FordMartin, Pacheco}.
An important motivation of Ref. \cite{Pacheco} was the precise calculation of
possible background events for the low-energy neutrino detector based on $^{115}$In.
Draxler et al. reported an indirect evidence for the "detour" transitions in the $\beta^-$
decay of $^{90}$Y, obtained by a polarization measurement \cite{Draxler}. In conclusion,
the virtual beta-gamma transitions have never been observed
in a direct and unambiguous way.

In this paper we present a clear evidence for the beta-gamma virtual transition
in the gamma spectrum of $^{59}$Ni. We start by introducing the correct theory
of the REC in case of \emph{2nu} forbidden transitions.
Then, we employ this theory to reanalyze experimental
results published by Janas et al. \cite{Janas}.
The difference between theoretical predictions for the direct \emph{2nu}
transition and the data is interpreted as coming
from the virtual transition through the excited state in $^{59}$Co. Furthermore,
we present the first complete theoretical description of the gamma spectrum for the specific
case of $^{59}$Ni which includes the REC part, the "detour" contribution as well
as the interference between these two channels.

\section{New model of REC}
Any search for the virtual transitions in EC decays
requires good understanding of the radiative capture, and this has a long
history completed only recently. The first approximation,
ignoring the Coulomb field of the nucleus and relativistic effects, valid only
for the K-capture, was provided by Morrison and Schiff in 1940 \cite{MS}. The model was
greatly advanced by Glauber and Martin \cite{GM,MG} who described exactly the
electron propagation in the Coulomb field, included relativistic and screening
effects, and considered captures from higher shells. This model was found to
be in good agreement with a large number of experiments \cite{Bambynek}. Nevertheless,
it is valid only for allowed decays (spin change $\Delta J = 0,1$, no parity change).
The extension of the REC theory to any order of forbiddenness, with inclusion
of Coulomb and relativistic corrections, was done by Zon and
Rapoport \cite{ZonRap,Zon}.
However, measurements of the REC spectra for first-forbidden
unique (\emph{1u}, $\Delta J = 2$, parity changes) K-captures in $^{41}$Ca and $^{204}$Tl
revealed strong discrepancies with their model \cite{41Ca,204Tl},
indicating that some approximations assumed by them are not valid.
The \emph{1u} decay is special as the probability ratio of radiative
to non-radiative electron capture does not depend on nuclear matrix elements.
The experiment delivers the absolute value of this ratio as a function of the
photon energy, which is compared without any normalization to the
theoretical prediction which does not have any adjustable parameters.
Kalinowski et al. \cite{Kalinowski1,Kalinowski2} made an attempt
to extend the Zon and Rapoport model by adding the "detour" transitions via virtual
nuclear states following ideas of Ford and Martin \cite{FordMartin}, but repeated
the same invalid approximations as Zon and Rapoport.
He was able to reproduce the spectrum of $^{41}$Ca at the cost of adding a
phenomenological parameter \cite{Kalinowski1}, but his model
failed in case of $^{204}$Tl \cite{204Tl}.
Finally, this conundrum was resolved by Pachucki et al. \cite{Pachucki} with a crucial insight
that the description of the electromagnetic radiation in this case is much more advantageous
in the \emph{length} gauge in contrast to the Coulomb gauge used in all previous
calculations. It allowed to expose and to circumvent
divergent terms which were the source of wrong approximations in the model
of Zon and Rapoport for the $\Delta J = 2$ transitions.
The problem was caused by neglecting the radiation from the nucleus
which is not justified in the Coulomb gauge.
The \emph{length}-gauge model, while
reproducing exactly the results of Glauber and Martin for allowed decays, was found
to describe very well both the shape and the intensity of experimental spectra
of $^{41}$Ca and $^{204}$Tl \cite{Pachucki} and later also the spectrum of
\emph{1u} K-capture in $^{81}$Kr \cite{81Kr}. It is because the
nuclear radiation is strongly suppressed in this gauge and can be neglected.

Following the method described in Ref. \cite{Pachucki}, here we extend the \emph{length}-gauge
model of REC to the \emph{2nu} transitions ($\Delta J = 2$, no parity change).
Conventionally, the emission probability of a photon with energy $k$
per ordinary non-radiative electron capture and per unit energy is expressed in the
form:
\begin{equation}
  \frac{1}{W_0}\frac{dW_R(k)}{dk} \equiv \frac{\alpha}{\pi\,m^2}\,\biggl(1-\frac{k}{Q}\biggr)^2\,k\,{\cal R}\,,
\end{equation}
where $\alpha$ is the fine structure constant, $m$ is the electron mass, $Q$ is the maximal photon
energy, and the units $\hbar=c=1$ are used. Details of the spectrum are thus represented
by the dimensionless shape factor $\cal R$. We note that in the simplest approximation
of Morrison and Schiff ${\cal R} = 1$ \cite{Bambynek}. For \emph{2nu} K-capture,
the \emph{length}-gauge model leads to:
\begin{equation}\label{}
  {\cal R}_{2nu} = \biggl(1-\frac{k}{Q}\biggr)^2\,{\cal R}^{(1)}(k) +
   \Lambda \biggl(\frac{k}{Q}\biggr)^2 \,{\cal R}^{(2)}(k)\,,
\end{equation}
where the functions ${\cal R}^{(1)}$ and ${\cal R}^{(2)}$, which incorporate all
Coulomb and relativistic corrections, are derived in Ref. \cite{Pachucki}.
In this case the spectrum depends on the parameter $\Lambda$ which is a combination of nuclear matrix elements relevant for the direct transition
between the initial and the final state:
\begin{equation}\label{}
  \Lambda \equiv \frac{|{\cal M}_{2nu}^{(4)}|^2}{|{\cal M}_{2nu}^{(2)}|^2} \,,
\end{equation}
with
\begin{equation}\label{}
  {\cal M}_{2nu}^{(n)} = \langle J_f \| r [ T_{211}-\frac{Z \alpha}{n\sqrt{5}}\bigl(\sqrt{2} T_{220}-\lambda \sqrt{3}\gamma^5 T_{221}\bigr)]  \| J_i \rangle \,,
\end{equation}
where $Z$ is the atomic number of the initial atom, $J_i$ and $J_f$ denote the initial, and the final
state, respectively and $T_{JLS}$ is a spherical tensor operator.
Eq.(2) was derived under the approximation $Z \alpha \ll 1$ and essentially represents the
photons emitted by the electron in the IB process. The nuclear contribution is strongly
suppressed in this model but \emph{only} when there are no close-lying excited states
in the daughter or in the parent nucleus \cite{Pachucki}.
We note that for $\Lambda = 1$ we obtain the shape factor for the K-capture
in \emph{1u} transition, which reproduced so well the experimental data \cite{Pachucki,81Kr}.
In addition, the shape factor ${\cal R}_{2nu}$, Eq.(2-4), has
the same form as in the model of Zon and Rapoport \cite{ZonRap} --- the important difference
is in the formula for the function ${\cal R}^{(2)}$ which represents electronic transitions
through $P_{3/2}$ and $D_{3/2}$ intermediate states. These transitions enter the description
of REC for nuclear decays with $\Delta J = 2$. The values of this function in our model,
and also of its Coulomb-free limit, differ significantly from the prediction of
Zon and Rapoport \cite{Pachucki}. This is the main reason why all previous attempts to identify
virtual transitions in the decay of $^{59}$Ni, which used either the wrong REC model
or its inaccurate Coulomb-free approximation, could not have been successful.

\section{Comparison with experiment}
To test the new REC prediction for \emph{2nu} decays,
we compare it to experimental data published for $^{59}$Ni \cite{Janas}.
The K-capture component of the REC spectrum was determined in Ref. \cite{Janas}
by recording gamma rays in coincidence with the KX-rays of cobalt.
The shape factors extracted from two independent measurements, using different detection
set-ups, are shown in Figure 2. The differences between two data sets reflect inaccuracies of
experimental procedures which involved various corrections, the largest being for
the Compton scattering and for the absolute efficiency of gamma detectors.
The prediction of the model is given by Eq.(2) where the functions
${\cal R}^{(1)}$ and ${\cal R}^{(2)}$ were calculated numerically for the case of $^{59}$Ni.
Since it is difficult to calculate the value of $\Lambda$, it is
treated as a free parameter of the model. We found that no value of this parameter
can reproduce both the shape and the intensity of the measured shape factor.
However, we can adjust this parameter to fit the low-energy part of the
spectrum. The least squares method for the points from both data sets with $k<600$~keV,
yields $\Lambda = 1.47\pm0.15$ and the resulting shape factor
is shown in Figure 2 by the dashed line.

\begin{figure}[htb]
\includegraphics[width=\columnwidth]{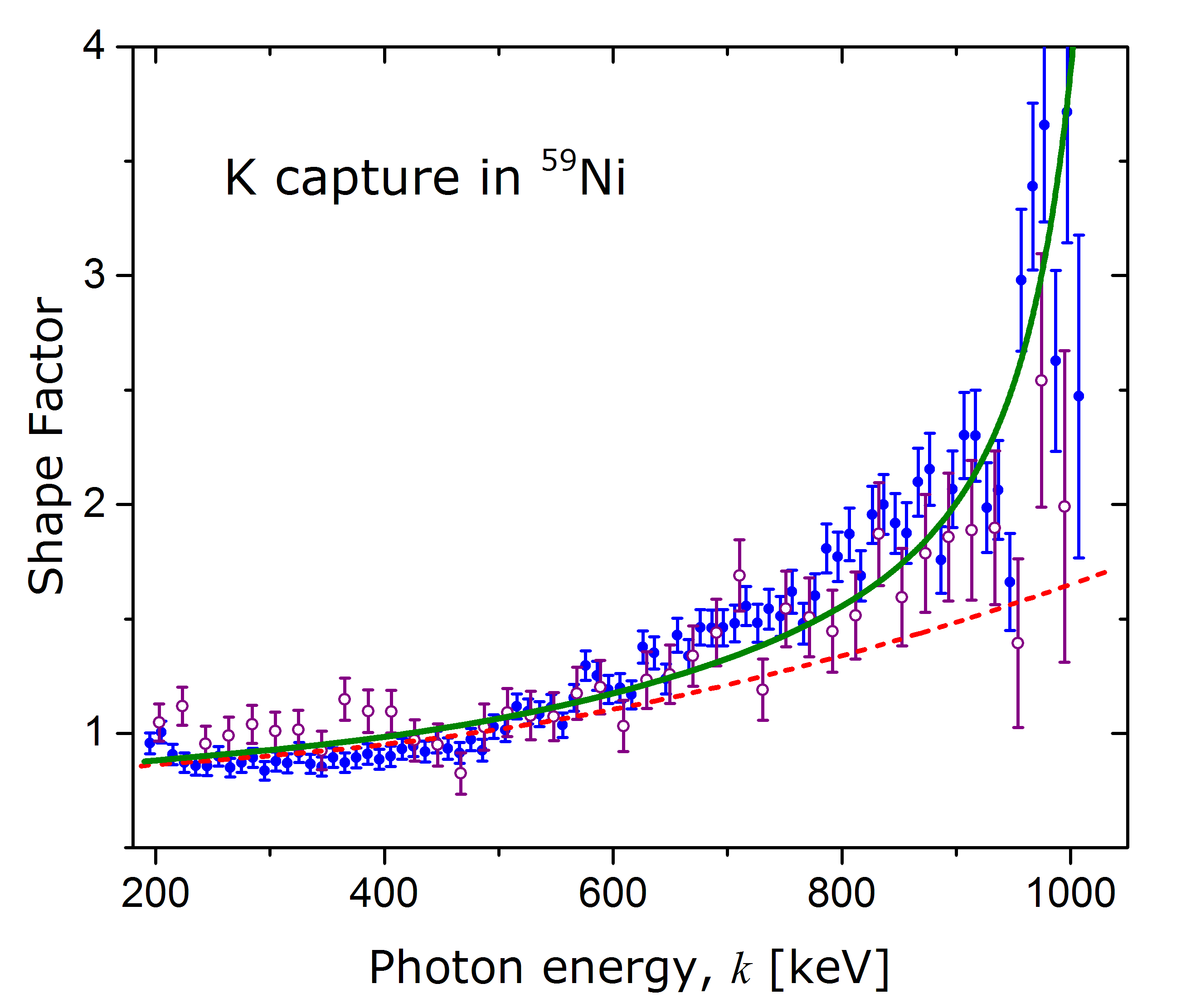}
\caption[T]{\label{fig:fig2}
(Color online) The radiative K-capture shape factor for $^{59}$Ni. Experimental points are from Ref. \cite{Janas}. The dashed line represents the REC prediction, Eq.(2),
for the value of the parameter $\Lambda = 1.47$. The solid line shows the best fitted full model, including the virtual "detour" transitions and the interference term.
}
\end{figure}

In Figure 2 we see that while the spectrum is well reproduced by the REC model
below 600~keV, the experimental intensity increases with energy
above the REC prediction, exactly as expected in the early papers of Rose et al. \cite{Rose}
and Lassila \cite{Lassila}. We interpret this excess of radiation at the
high-energy end as the evidence for the virtual beta-gamma transition through the
excited $3/2^-$ level in $^{59}$Co.

\section{Virtual transitions}
To verify this hypothesis, we calculate the
spectrum of photons emitted in the process of virtual transition, marked in Figure 1 by
dashed arrows, using the \emph{length}-gauge approach.
As the result, two additional terms have to be added to the shape factor:
\begin{equation}\label{}
   {\cal R}_{tot} =  {\cal R}_{2nu} + {\cal R}_{vir} \pm {\cal R}_{int} \,,
\end{equation}
where the contribution from the virtual transition is given by:
\begin{equation}\label{}
  {\cal R}_{vir}= \frac{27 \pi}{450} \, \frac{m^2 \, k^4}{Q^2 (E^*-k)^2}\,
  \frac{{\cal M}_{E2}^2 ({\cal M}_F^2 + {\cal M}_{GT}^2) }{|{\cal M}_{2nu}^{(2)}|^2} \,,
\end{equation}
and the interference term reads:
\begin{equation}\label{}
  {\cal R}_{int} = \frac{9 \sqrt{2 \pi}}{25} \, \frac{m \,k^2}{E^*-k} \,
  \frac{f(E2)}{R \, Q^2} \, \sqrt{\Lambda} \,
  \frac{{\cal M}_{E2} (\sqrt{\frac{5}{3}}{\cal M}_F + {\cal M}_{GT}) }{{\cal M}_{2nu}^{(2)}} \,.
\end{equation}
$E^*$ is the excitation energy of the $3/2^-$ level in $^{59}$Co, $R$ is the nuclear radius,
and $f(E2)$ is expressed in terms of functions defined in Ref. \cite{Pachucki}:
\begin{equation}\label{}
  f(E2) = \frac{F^<_2(R)}{{\cal G}(R)} \, \sqrt{{\cal R}_{E2}(S_{1/2}\rightarrow D^>_{3/2})} \,.
\end{equation}
The additional nuclear matrix elements which enter the model refer to the electromagnetic E2
transition and to Fermi and Gamow-Teller beta transitions:
\begin{eqnarray}
  {\cal M}_{E2} &=& \langle J_f \| r^2 \, T_{220} \| J_{f'} \rangle \,, \\
  {\cal M}_{F}  &=& \langle J_{f'} \|  T_{000} \| J_i \rangle \,,\nonumber \\
  {\cal M}_{GT} &=& \langle J_{f'} \|  \lambda \, \gamma^5 \, T_{101} \| J_i \rangle \nonumber \,,
\end{eqnarray}
where $J_{f'}$ represents the intermediate $3/2^-$ state.

The formulas above may be simplified in the case of $^{59}$Ni by noting that the matrix element
for the Fermi transition is negligible. The Fermi transition in question can
proceed only due to admixtures of the $T=5/2$ level, being the isospin analog of the $3/2^-$ state in
$^{59}$Co, to the ground state of $^{59}$Ni. This analog level is located in $^{59}$Ni
at about 8.4~MeV excitation energy \cite{NNDC}, which makes the isospin
admixtures very small. Following Ref. \cite{Bhatt} we can expect the Fermi
matrix element to be of the order of $10^{-3}$.
Therefore, we put ${\cal M}_F = 0$ and
we see that the both new terms (Eqs. 6,7) depend on the single additional combination of
nuclear matrix elements:
\begin{equation}\label{}
  \chi \equiv \frac{{\cal M}_{E2} \, {\cal M}_{GT}}{{\cal M}_{2nu}^{(2)} \, \lambdabar } \,.
\end{equation}
where the Compton wavelength of the electron, $\lambdabar = 386.2$~fm, is inserted to ensure
that the $\chi$ is dimensionless.

Now, treating the $\Lambda$ and $\chi$ as free parameters we fit both data sets in the
full energy range. The least-squares minimum is reached for the values:
\begin{equation}\label{}
  \Lambda=1.52 \pm 0.18, \,\,\, \chi=0.31_{-0.34}^{+0.17} \,,
\end{equation}
and for the positive interference sign.
The result, shown in Figure 2 with the solid line, reproduces the shape factor very well.
The value of $\Lambda$ best fitting the full spectrum is almost equal to the
value determined from the analysis of the low-energy part with the pure REC model.
This validates our assumption that the virtual contribution does not affect
the spectrum at low energy. Figure 3 shows the contour plot of the least-squares
function. Due to its rather flat minimum, the values of both parameters cannot
be extracted precisely.
The resulting total gamma emission probability, given by integration of Eq.(1) for $k>195$~keV,
according to prediction, is $7.92 \times 10^{-4}$.
The integrated virtual and interference components represent 2\% and 2.3\% of this
intensity, respectively.

All matrix elements entering this analysis can be estimated from
available experimental information.
First, we note that the value of $\Lambda$ can be determined from the
$\beta^+/K$ ratio for the decay of $^{59}$Ni. This ratio was measured
in Ref. \cite{Janas} and yielded the value $\Lambda = 1.0 \pm 0.3$
which agrees within $2\sigma$ with the value determined in the present analysis.
From the half-life of $^{59}$Ni, using the K-capture rate formula given in
Ref. \cite{Pachucki}, we determine that $|{\cal M}_{2nu}^{(2)}| = 0.0349$ fm.
The $3/2^-$ level in $^{59}$Co at 1099.3 keV has a half-life of $3.1(4)$~ps \cite{NNDC}.
From this value the E2 transition matrix element can be calculated
yielding $|{\cal M}_{E2}|=21.3$~fm$^2$. The ${\cal M}_{GT}$ element
can be estimated from the systematics of the reduced half-life values,
$ft$, for the Gamow-Teller transitions in the region of $^{56}$Ni \cite{NNDC}.
The measured values suggest that in our case $\log ft = 6$ is a reasonable assumption.
Using this value and the relation \cite{TownerHardy}:
\begin{equation}\label{}
f t = \frac{6144 \,{\rm s}}{B_{GT}}, \,\,\,\, B_{GT}=\frac{1}{2 J_i+1}{\cal M}_{GT}^2
\end{equation}
we arrive at $|{\cal M}_{GT}|=0.157$.
Inserting these values in Eq.(10) leads to $|\chi| = 0.25$ which agrees
well with the value determined from the spectrum of $^{59}$Ni.

In principle, all these matrix elements can be calculated within
the nuclear structure theory. This challenging task remains outside
the scope of this paper. We note, however, that those nuclear models which
offer precise predictions of spectroscopic quality, like the shell-model,
usually employ effective or empirical nucleon-nucleon interactions.
The theoretical results derived in Ref. \cite{Pachucki} and here,
based on elementary quantum electrodynamics and weak interaction
theory, do not involve any such phenomenological ingredients.
Thus, the combinations of matrix elements, $\Lambda$ and $\chi$, which
can be measured with higher precision in the future, in fact can be used to constrain
models of nuclear structure.

\begin{figure}[htb]
\includegraphics[width=\columnwidth]{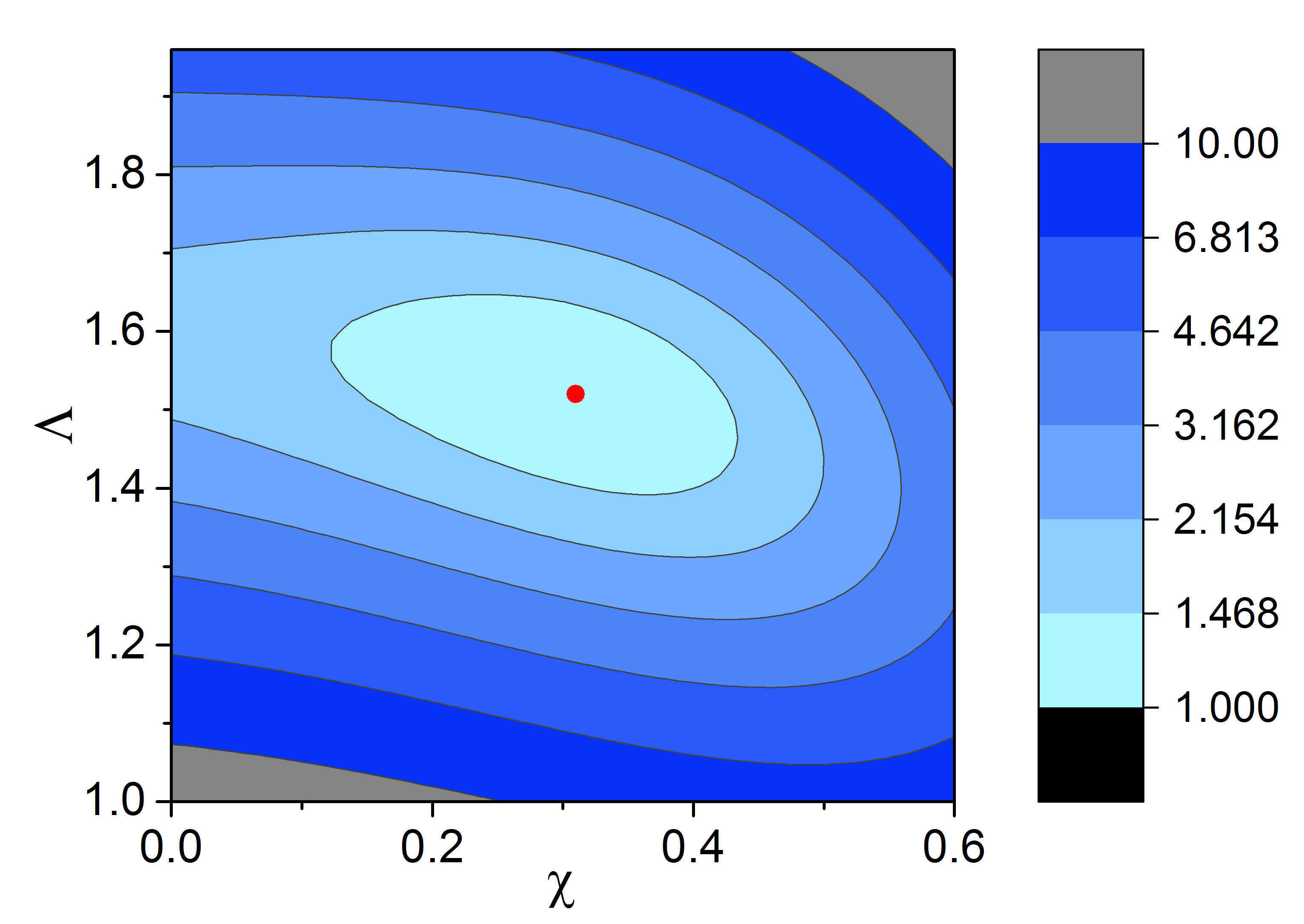}
\caption[T]{\label{fig:fig4}
(Color online) The contour map of the minimizing function.
The dot shows the position of the minimum.
}
\end{figure}


\section{Summary}
The electromagnetic radiation accompanying the nuclear EC decay
has two components: the electronic and the nuclear. The absolute discrimination between them
is meaningless as their relative contribution depends on the adopted gauge of the
electromagnetic field. Recently, a long-standing puzzle of REC in \emph{1u}
transitions was resolved by a new theoretical method which utilized the \emph{length}-gauge
which suppresses the nuclear part and thus allows one to neglect it \cite{Pachucki}.
This suppression, however, is not complete when there are states in the daughter
nucleus close to the initial state. Then, an additional contribution of these states
to the emitted radiation must be taken into account. We find such a situation in
$^{59}$Ni which decays by \emph{2nu} EC to $^{59}$Co. In the daughter
nucleus there exist an excited state located only 26~keV above the initial state.
First, we applied the new model to \emph{2nu}
transitions and compared it with the experimental data published for the decay of $^{59}$Ni.
This comparison revealed an excess of radiation in the high energy end of the gamma spectrum.
We claim that it originates from virtual transitions through this excited state
in $^{59}$Co. Second, we derived the complete model, which includes the contribution from the
virtual process, composed of allowed EC and E2 gamma transitions and its interference
with the REC effect. It depends on two parameters involving nuclear matrix elements.
We showed that when these parameters have values consistent with the estimated
values of the relevant matrix elements, the model describes very well
the measured spectrum. This finding is the first direct and unambiguous observation of virtual
beta-gamma transitions. The radiation emitted in the EC decay of $^{59}$Ni
is the coherent sum of both electronic and nuclear components.
The latter is responsible for about 4\% of the total intensity.



%

\end{document}